\documentclass[preprint,times]{elsarticle}

\usepackage[margin=1in]{geometry}

\usepackage{graphicx}
\usepackage{amssymb}
\usepackage{enumerate}
\usepackage{float}
\usepackage{hhline}
\usepackage{multirow}
\usepackage{url}
\usepackage[linesnumbered,ruled,vlined]{algorithm2e}

\usepackage{setspace}
\usepackage{lineno}
\usepackage[absolute,overlay]{textpos}

\usepackage{fancyhdr}
 
\newcommand{\Eqn}[1]{(#1)} 
\newcommand{\Fig}[1]{figure #1} 
\newcommand{\Tab}[1]{table #1} 

\begin{document}

\doublespacing

\title{An inhomogeneous most likely path formalism for proton computed tomography}

\author[add1,add2]{Mark D. Brooke}
\ead{mark.brooke@oncology.ox.ac.uk}
\author[add2,add3]{Scott N. Penfold}

\address[add1]{Department of Oncology, University of Oxford, Old Road Campus, Roosevelt Drive, Oxford, OX3 7DQ, United Kingdom}
\address[add2]{Department of Physics, University of Adelaide, Adelaide, South Australia, 5005, Australia}
\address[add3]{Department of Medical Physics, Royal Adelaide Hospital, Adelaide, South Australia, 5000, Australia}

\begin{abstract}
\textbf{Purpose:} Multiple Coulomb scattering (MCS) poses a challenge in proton CT (pCT) image reconstruction. The assumption of straight paths is replaced with Bayesian models of the most likely path (MLP). Current MLP-based pCT reconstruction approaches assume a water scattering environment. We propose an MLP formalism based on accurate determination of scattering moments in inhomogeneous media.

\noindent\textbf{Methods:} Scattering power relative to water (RScP) was calculated for a range of human tissues and investigated against relative stopping power (RStP). Monte Carlo simulation was used to compare the new inhomogeneous MLP formalism to the water approach in a slab geometry and a human head phantom. An MLP-Spline-Hybrid method was investigated for improved computational efficiency.

\noindent\textbf{Results:} A piecewise-linear correlation between RStP and RScP was shown, which may assist in iterative pCT reconstruction. The inhomogeneous formalism predicted Monte Carlo proton paths through a water cube with thick bone inserts to within 1.0 mm for beams ranging from 210 to 230 MeV incident energy. Improvement in accuracy over the conventional MLP ranged from 5\% for a 230 MeV beam to 17\% for 210 MeV. There was no noticeable gain in accuracy when predicting 200 MeV proton paths through a clinically relevant human head phantom. The MLP-Spline-Hybrid method reduced computation time by half while suffering negligible loss of accuracy.

\noindent\textbf{Conclusions:} We have presented an MLP formalism that accounts for material composition. In most clinical cases a water scattering environment can be assumed, however in certain cases of significant heterogeneity the proposed algorithm may improve proton path estimation.

\end{abstract}

\maketitle

\begin{textblock*}{20cm}(1cm,1cm) 
   \footnotesize \includegraphics[height=12px]{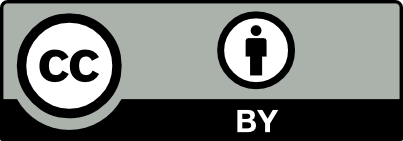} \textcopyright \ 2020. This manuscript version is made available under the CC-BY 4.0 license https://creativecommons.org/licenses/by/4.0/
\end{textblock*}

\section{Introduction}

The main advantage of proton therapy over conventional X-ray radiation therapy is the \emph{Bragg peak} in energy deposition, a phenomenon whereby most of a heavy charged particle's energy per unit length is deposited toward the end of its trajectory in a sharp peak. This quality is advantageous in radiotherapy treatment using such particles as beam energy and intensity can be manipulated to deposit a highly conformable dose to the tumor volume, with a low dose on entry and no exit dose.

The range of protons in matter is dependent on the stopping power of the material through which it travels. To accurately predict the location of the Bragg peak, therefore, accurate knowledge of the stopping power of the patient tissues is required. Tissues can be characterised by their \emph{relative stopping power} (RStP). This is simply the stopping power of a material relative to that of water at a given energy. For human tissues, the RStP is approximately independent of proton energy for therapeutic energies.

In current clinical practice, RStP is estimated by conversion of X-ray CT Hounsfield units via an empirically derived calibration curve \citep{schneideretal}. This approach can lead to errors in stopping power of up to 3\% \citep{smith,jiang}. Proton CT (pCT) is an alternative approach in which RStP of the patient is measured directly with an energetic proton beam.

In conventional CT imaging, X-rays are assumed to move in straight lines, whereas protons are known to undergo multiple Coulomb scattering (MCS), which poses a challenge for image reconstruction that is absent from X-ray CT imaging. In order to perform a pCT reconstruction, knowledge of the position, direction and energy of individual protons before and after they traverse the patient is required. Hanson \emph{et al.} first proposed the concept of estimating proton paths by measuring their exit position post-patient \citep{Hanson78,Hanson79,Hanson81}. In the pCT system concept and design by Schulte \textit{et al.} in 2004 \citep{schulte}, measurements of individual proton positions and directions pre- and post-patient are obtained through two 2-dimension sensitive tracking modules upstream and downstream of the patient. Each tracking module consists of 
orthogonally oriented single sided silicon strip detectors (SiSD) that are position-sensitive in one dimension. The energy of an incoming proton is assumed to be equal to the energy at which it is ejected from the accelerator, while the residual energy post-patient is measured by a segmented scintillation crystal calorimeter.

Assuming a straight line path in proton radiography leads to poor spatial resolution \citep{schneiderpedroni}, so it is therefore necessary to develop a probabilistic model for the movement of protons through matter, known as a \emph{most likely path} (MLP) formalism. In this paper we use the matrix-based MLP formula proposed by Schulte \emph{et al.} \citep{penfoldMLP}, which followed the methods of Williams \citep{williams}. In the Fermi-Eyges framework, Schulte \emph{et al.} have applied Bayesian techniques to a bivariate Gaussian distribution to calculate the most likely lateral position and angular deflection at an intermediate depth, given the entry and exit conditions of the proton. The distribution provides an error matrix given by equation (27) in their paper, from which the lateral error can be calculated at any depth in the medium to define a probability envelope surrounding the most likely path. Typically in pCT iterative reconstruction, the (deterministic) path of a proton is mapped out by assigning binary values to each voxel; 1 if it passes through that voxel, and 0 otherwise. Using a probability envelope surrounding the MLP, it is possible to assign to each voxel a continuous value between 0 and 1 which represents the probability of the proton traversing that voxel. Wang \textit{et al}. \citep{wangmackietome} demonstrated that using a 90\% probability map directly for pCT reconstruction can yield a smoother reconstructed image when compared to using deterministic (binary) path maps. It is worth noting that the computation time in the method proposed by Wang \textit{et al}. \citep{wangmackietome} is increased significantly due to calculations involving non-sparse matrices. 

The incompleteness of the formalism by Schulte \emph{et al.} \citep{penfoldMLP} stems from the fact that the compact MLP formula is derived for homogeneous media, with the covariance matrices calculated assuming all scatter takes place in water. Such an approximation may lead to inaccuracies in the proton path estimate during pCT reconstruction and result in substandard spatial resolution in the reconstructed image. A study by Wong \emph{et al.} investigated the effect of tissue inhomogeneity on the accuracy of the water-based MLP estimate \citep{Wong2009}. They found the error in proton path reconstruction to be up to 20\% larger in a 20 cm slab geometry which included small bone inserts and air cavities. There was no attempt, however, to adjust the MLP formalism to account for these materials. 

In this paper, we propose an inhomogeneous MLP formalism, based on the ratio of scattering power in the material of interest to that in water. We refer to this quantity as the \emph{relative scattering power} (RScP) hereon. Using the scattering power formulae of Gottschalk \citep{gottschalkRadioProtons}, the RScP at any depth is dependent only on the material at that depth, even when considering nonlocal effects of MCS buildup in the material preceeding the calculation point. It is therefore straightforward to calculate the Fermi-Eyges scattering moments \citep{Eyges1948} in inhomogeneous media when the composition is known \emph{a priori}. Furthermore, we are able to investigate a relationship between RScP and RStP in different materials. In a 2017 study, Collins-Fekete \emph{et al.} \citep{Collins2017} presented an inhomogeneous MLP formalism in which the Fermi-Eyges moments are calculated using a method proposed by Kanematsu \citep{Kanematsu2008}, however this requires integrating a material dependent quantity, the radiation length, over depth. Gottschalk's formulae are both more accurate and better suited to mixed slab geometries than the differential Highland formula used by Kanematsu \citep{gottschalkRadioProtons}. 

In an attempt to improve computational efficiency, several methods using cubic spline trajectories have been proposed as an alternative to Bayesian calculation of the MLP \citep{tianfang,wang2011,feketeCSP}. In this work, we introduce an inhomogeneous MLP-Spline-Hybrid method which involves the use of cubic splines fit through a limited number of Bayesian MLP points.

We have used the \emph{TOPAS} tool \citep{TOPAS}, which wraps and extends the \emph{GEANT4} simulation toolkit \citep{geant4}, to compare our new inhomgeneous MLP formalism to that of Schulte \emph{et al.} \citep{penfoldMLP}. We show a piecewise-linear relationship between RStP and RScP for a range of human tissues and describe how this relationship may be implemented in iterative pCT reconstruction.

\section{Methods}

In the compact matrix-based MLP formula of Schulte 
\emph{et al.} \citep{penfoldMLP} the most likely lateral position $t_1$ and angular deflection $\theta_1$ at an intermediate depth $u_1$ are represented by the vector $\mathbf{y}_1 = \left( t_1 \ \theta_1 \right)^T$,
given the respective entry and exit conditions,
$\mathbf{y}_{\mathrm{in}} \equiv \mathbf{y}_0 = \left( t_0 \ \theta_0 \right)^T$ at depth $u_0$, and $\mathbf{y}_{\mathrm{out}} \equiv \mathbf{y}_2 = \left( t_2 \ \theta_2 \right)^T$ at depth $u_2$.
The MLP is calculated by Equation (24) in their work \citep{penfoldMLP},
\begin{equation}
\mathbf{y}_{\mathrm{MLP}} = \left( \Sigma_1^{-1} + R_1^T\Sigma_{2}^{-1}R_1 \right)^{-1}\left( \Sigma_1^{-1}R_{0}\mathbf{y}_{0} + R_1^T\Sigma_{2}^{-1}\mathbf{y}_{2} \right) \label{mlp}
\end{equation}
where $R_0$ and $R_1$ are change-of-basis matrices,
\begin{equation}
R_0 = 
\left(
\begin{array}{c c}
1 & u_{1} - u_0 \\
0 & 1
\end{array}
\right), \ \ \ 
R_1 = 
\left(
\begin{array}{c c}
1 & u_{2} - u_1 \\
0 & 1
\end{array}
\right), \label{changeofbasismatrices}
\end{equation}
and $\Sigma_1$ and $\Sigma_2$ are the covariance matrices,
\begin{equation}
\Sigma_1 = 
\left(
\begin{array}{c c}
\sigma^2_{t_1} & \sigma^2_{t_1\theta_1} \vspace{2px} \\
\sigma^2_{t_1\theta_1} & \sigma^2_{\theta_1} 
\end{array}
\right), \ \ \
\Sigma_2 = 
\left(
\begin{array}{c c}
\sigma^2_{t_2} & \sigma^2_{t_2\theta_2} \vspace{2px} \\
\sigma^2_{t_2\theta_2} & \sigma^2_{\theta_2} 
\end{array}
\right). \label{covmatrices}
\end{equation}
The elements of the covariance matrices in \Eqn{\ref{covmatrices}}, known as the \emph{scattering moments}, are given by (for $i=1,2$) \citep{Eyges1948}
\begin{eqnarray}
\sigma_{\theta_i}^2 \equiv A_0(u_i) &=& \int_{u_{i-1}}^{u_i} T(\eta)d\eta, \label{scatmomentsSigma_1}\\
\sigma_{t_i\theta_i}^2 \equiv A_1(u_i) &=& \int_{u_{i-1}}^{u_i} (u_i-\eta)T(\eta)d\eta, \label{scatmomentsSigma_2}\\
\sigma_{t_i}^2 \equiv A_2(u_i) &=& \int_{u_{i-1}}^{u_i} (u_i-\eta)^2 T(\eta)d\eta, \label{scatmomentsSigma_3}
\end{eqnarray}
where $T(u)$ is the \emph{scattering power}, which is defined as the rate of increase, with depth $u$, of the mean square of the projected scattering angle $\theta$. That is, 
\begin{equation} 
T(u) \equiv \frac{d\langle \theta^2 \rangle}{du}.\label{scatpower1}
\end{equation}

The variance in lateral displacement is given by the (1,1) matrix element of \Eqn{\ref{errormatrixexact}};
\begin{equation}
 \epsilon_{t_1\theta_1} = 2\left(\Sigma_1^{-1} + R_1^T\Sigma_2^{-1}R_1\right)^{-1}. \label{errormatrixexact}
\end{equation}
This variance reflects the square of the standard error in the MLP and may be used to define a probability envelope surrounding the most likely path. This probability envelope is defined such that it contains at least a specified proportion of all possible MCS influenced paths. For example, given a pair of entry and exit information, only 0.3\% of all possible particle trajectories should exit a $3\sigma$ envelope surrounding the MLP. Note that \Eqn{\ref{errormatrixexact}}, and therefore the size of the envelope, is dependent on the material composition of the medium traversed by the particle.

\subsection{Scattering power calculations for inhomogeneous materials}\label{scatteringpowersection}

Here, we present a method for directly calculating the depth dependent scattering power $T(u)$, and hence the scattering moments, \Eqn{\ref{scatmomentsSigma_1}} to \Eqn{\ref{scatmomentsSigma_3}}, in various materials. This method does not require the empirical approximations from Lynch and Dahl \citep{lynchdahl} that are used in the homogeneous formalism of Schulte \emph{et al.} \citep{penfoldMLP}.

In the Fermi-Eyges framework, in which a Gaussian scattering process is assumed, scattering power may be calculated from first principles using the scattering cross-section. Using the cross-section formula of Goudsmit and Saunderson \citep{gs1,gs2}, which is better behaved at small angles than the well-known Mott \citep{mott} and Rutherford \citep{Rutherford1911} formulae, Gottschalk derived an expression for scattering power, which could be considerably simplified for radiotherapy protons in the energy range of 3 to 300 MeV \citep{gottschalkRadioProtons}, by first defining a quantity, he termed the \emph{scattering length}, $1/X_s$. For a scattering medium with mass density $\rho$, atomic number $Z$, and mass number $A$, the scattering length is defined as
\begin{equation}
\frac{1}{X_s} \equiv \alpha N_A \rho \left( \frac{e^2}{m_ec^2} \right)^2\frac{Z^2}{A}\left\{ 2\ln\left[ 33219(AZ)^{-1/3}\right]-1\right\}. \label{scatteringlength}
\end{equation}
Here, $\alpha$ is the fine structure constant, $N_A$ is Avogadro's number, $e$ is the charge of an electron, and $m_ec^2$ is the rest mass of an electron. Using this scattering length, it can be shown that the scattering power obeys
\begin{equation}
T(u) = \frac{2\pi}{\alpha}\left( m_ec^2 \right)^2\left( \frac{\tau+1}{\tau+2} \right)^2\frac{1}{E^2(u)}\frac{1}{X_s}  \label{gottschalkscatpower2}
\end{equation}
where $E(u)$ is the depth-dependent kinetic energy of the proton and its \emph{reduced kinetic energy}, $\tau$, is defined as
\begin{equation}
\tau \equiv \frac{E(u)}{m_pc^2} \label{reducedkinenergy}
\end{equation}
where $m_pc^2$ is the rest mass of a proton. In addition to Gottschalk's work \citep{gottschalkRadioProtons}, the reader is referred to Rossi's book \citep{rossibook} for clarification of the underlying theory leading to this result.

In order to calculate the scattering power for composite materials, such as those found in the human body, we notice that the only material dependent part of \Eqn{\ref{gottschalkscatpower2}} is the scattering length. Thus, if the composite material consists of $n$ elements, each with fractional weight per volume $0<w_k\leq 1, k=1,\dots,n$ then
\begin{equation}
\frac{1}{X_s} = \rho \sum_{k=1}^n w_k \left( \frac{1}{\rho X_s} \right)_k \label{scatlengthsum}
\end{equation}
where $\rho$ is the mass density of the composite material.

We wish to catalogue materials based on a single energy-independent value, analagous to the RStP, which can be assumed valid at any depth. This can be achieved by defining the \emph{relative scattering power} (RScP), $\hat{T}$, as the ratio of the scattering power for the material to the scattering power for water at the same energy. Using Gottschalk's formula \Eqn{\ref{gottschalkscatpower2}}, the energy dependent terms cancel and the result is simply the ratio of the scattering lengths. It can be shown that
\begin{equation}
\hat{T} = 0.01636 \ \rho\sum_{k=1}^n w_k \frac{Z^2}{A}\left\{ 2\ln\left[ 33219(AZ)^{-1/3}\right]-1\right\} \label{relscatpower}
\end{equation}
with weights $w_k$ and mass density $\rho$ (in units of g/cm$^3$) as defined in (\ref{scatlengthsum}). The factor of 0.01636 is the reciprocal of the summation term applied to liquid water. In practice, we can now calculate the scattering power at any depth, and hence the covariance matrices in \Eqn{\ref{covmatrices}} using
\begin{equation}
T(u) = T_\mathrm{w}(u)\hat{T} \label{totalscatpower}
\end{equation}
where the subscript 'w' refers to the value for water.

\subsection{Stopping power calculations for inhomogeneous materials}\label{stoppingpowersection}

If we wish to calculate scattering power, and hence the scattering moments, using \Eqn{\ref{totalscatpower}} then we must have an estimate of the kinetic energy of the proton at depth $u$. This can be calculated through an appropriately weighted combination of the forward and backward Euler methods, using stopping powers. 

One can rearrange the definition of stopping power,
\begin{equation}
S(E) \equiv -\frac{dE}{du}, \label{stoppingpower}
\end{equation}
to find the energy lost per unit length if the stopping power is known. In discretizing the depth we have $u_j$ for $j=0,\dots,N$ for some positive integer $N$. The spacing between successive depths is constant and equal to $\delta u = u_j-u_{j-1}$ for $j=1,\dots,N$. The forward Euler method gives
\begin{equation}
E_j^F = E_{j-1}^F - S(E_{j-1}^F)\delta u, \ \ \ j=1,\dots,N \label{eulerforward}
\end{equation}
with a boundary condition of $E_0^F = E_\mathrm{in}$, the incoming energy of the proton as it enters the medium, assumed to be equal to the exiting energy from the accelerator. The backward Euler method gives
\begin{equation}
E_j^B = E_{j+1}^B + S(E_{j+1}^B)\delta u, \ \ \ j=0,\dots,N-1 \label{eulerbackward}
\end{equation}
with a boundary condition of $E_N^F = E_\mathrm{out}$, the outgoing energy of the proton as it exits the medium, as measured by a detector.

Errors will accumulate with successive steps in each method. Hence it is reasonable to perform a weighted average of both methods. For example, if more steps must be taken in the backward method than in the forward, the forward method is expected to be more accurate and will be weighted more heavily. $E_j^F$ is multiplied by the normalised length between $u_j$ and $u_N$, while $E^B$ is multiplied by the normalised length between $u_0$ and $u_j$. That is,
\begin{equation}
E_j = \frac{N-j}{N}E_j^F + \frac{j}{N}E_j^B, \ \ \ j=1,\dots,N-1 \label{eulercombo}
\end{equation}
with boundary conditions $E_0=E_\mathrm{in}$ and $E_N = E_\mathrm{out}$.

Stopping powers can be determined using the Bethe-Bloch formula \citep{betheashkin},
\begin{equation}
S(E) \equiv -\left\langle \frac{dE}{du} \right\rangle = \frac{4\pi}{m_ec^2}\frac{\rho_e}{\beta^2}\left( \frac{e^2}{4\pi\epsilon_0}\right)^2\left[ \ln\left( \frac{2m_ec^2\beta^2}{I(1-\beta^2)}\right)-\beta^2 \right] \label{bethebloch}
\end{equation}
where $\rho_e$ is the electron number density of the material, $I$ is the \emph{mean excitation potential}, $\epsilon_0$ is the vacuum permittivity, and $\beta \equiv v/c$ where $v$ is the speed of the proton and $c$ is the speed of light in a vacuum. Experimental values of the mean excitation potential for a single-element substance (e.g. O$_2$ gas) can be obtained from a database, but for more complicated composite materials consisting of $n$ elements, one may use the Bragg additive rule,
\begin{equation}
\ln I = \sum_{k=1}^n w_k \frac{Z_k}{A_k} \ln I_k \left\langle \frac{Z}{A} \right\rangle^{-1} \label{additiverule}
\end{equation}
where
\begin{equation}
\left\langle \frac{Z}{A} \right\rangle = \sum_{k=1}^n w_k \frac{Z_k}{A_k}.
\end{equation}
$w_k$ is the fractional weight per volume of the $k$-th element, as in \Eqn{\ref{scatlengthsum}}.

Similar to relative scattering power, we may define the \emph{relative stopping power} (RStP), $\hat{S}$, as the stopping power in the material, $S$, divided by the stopping power in water, $S_\mathrm{w}$, at the same energy. That is,
\begin{equation}
\hat{S} = \frac{S}{S_\mathrm{w}}. \label{relstoppingpower}
\end{equation}
Note that the RStP is weakly energy dependent, however the effect for most materials and energies encountered clinically can be assumed negligible. This is explored in more detail in section \ref{cataloguesec}. Now, if the material through which the proton is passing is known, then one may calculate the stopping power by simply multiplying the stopping power in water by the RStP for that material,
\begin{equation}
S(u) = S_\mathrm{w}(u)\hat{S}. \label{totalstoppower}
\end{equation}
In the proposed use of this method, the current estimate of RStP from the reconstructed pCT image will be used to approximate RStP in each voxel. The forward and backward Euler methods can now be alternatively written, respectively, as
\begin{eqnarray}
&& E_j^F = E_{j-1}^F - \hat{S}_{j-1}S_\mathrm{w}(E_{j-1}^F)\delta u, \ \ \ j=1,\dots,N, \label{eulerforward2} \\
&& E_j^B = E_{j+1}^B + \hat{S}_{j+1}S_\mathrm{w}(E_{j+1}^B)\delta u, \ \ \ j=0,\dots,N-1 \label{eulerbackward2}
\end{eqnarray}
where $\hat{S}_j$ is the relative stopping power at depth $u_j$.

The calculation of the MLP is implemented in a two-stage process, assuming there exists a catalogue of RScP and RStP values for all materials of interest (refer to section \ref{cataloguesec}). The first stage involves estimating the proton energy at discrete depths within the patient, using (\ref{eulercombo}), which is then used to calculate the material-dependent scattering power, using (\ref{totalscatpower}). The second stage is to calculate the scattering moments, (\ref{scatmomentsSigma_1}) to (\ref{scatmomentsSigma_3}), using numerical integration, and therefore calculate the most likely position at each discrete depth using (\ref{mlp}). More algorithmic details of our implementation can be found in section \ref{MCsec} and in the Appendix.

\subsection{MLP-Spline-Hybrid method}

In an effort to increase computational efficiency, the inhomogeneous MLP may be calculated using the aforementioned methods only at subsampled depths within the phantom. The entire trajectory is then estimated by fitting a cubic spline through these points, given the initial and final directions of the proton as measured by detectors placed pre- and post-patient. The upper and lower bounds of the probability envelope are constructed by fitting cubic splines through the points generated by adding and subtracting, respectively, the lateral error, using \Eqn{\ref{errormatrixexact}}, to the MLP estimate at the calculation points. A multiplier can be applied to the lateral error to set the significance level. For example, a multiplier of 3 creates a 3$\sigma$ envelope which should contain 99.7\% of possible proton trajectories for given entry and exit information.

We propose that the sampled points be chosen as distinct material boundaries, so that the number of samples used for fitting the spline scales with the complexity of the heterogeneity. These boundaries may be identified by thresholded RStP differences between air, tissue and bone. We refer the reader to section \ref{discussionsec} for a discussion on how this may be implemented in the context of iterative pCT reconstruction.

The method described above shall be referred to as the inhomogeneous MLP-Spline-Hybrid method and denoted MLP$_\mathrm{x}$SH hereon.

\subsection{Catalogue of anatomical materials}
\label{cataloguesec}

We have thus far explored methods for calculating the relative scattering and stopping powers in various materials. In iterative pCT reconstruction, one may first assume the imaged object is made purely of water, and on each successive iteration the internal composition may be updated. Prior to the reconstruction, the depth at which some material is located is unknown, and it is for this reason that we wish to characterise materials by energy independent RStP and RScP values (as proton energy is related to penetration depth). We have already observed that our expression for RScP in \Eqn{\ref{relscatpower}} is dependent only on material composition, and not on the proton energy, which is a direct result of using Gottschalk's derived scattering power \citep{gottschalkRadioProtons}. The validity of energy independent RStP is, however, not so straightforward. We explore this below.

RScP ($\hat{T}$) and RStP ($\hat{S}$) were calculated for the forty-six human tissues listed in ICRU Report 46 \citep{ICRU46}, which gave their elemental composition, mass density and electron number density. Values were also calculated for air, which has been approximated as consisting 79\% of Nitrogen and 21\% of Oxygen. ($\rho_e = 0.001225$ g/cm$^3$ has been used as the electron number density.) Mean excitation values used in the calculation of stopping power were obtained, for the same set of materials, from ICRU Report 49 \citep{ICRU49}. The nominated RStP assigned to each material was arbitrarily defined as the average of the energy dependent RStP$(E)$ calculated at 1 MeV incremeents over the range $10\leq E \leq 300$ MeV.

Arbor \emph{et al.} quote a region of negligible energy dependence between 80 and 300 MeV, by comparing \emph{GEANT4} RStP values at various proton energies to that at 300 MeV \citep{Arbor2015}. They used a clinical \emph{Gammex 467} tissue characterisation phantom, for which the material contributing to the greatest RStP variation (approximately 1\%) was the cortical bone equivalent, \emph{Gammex SB3}. Our calculations show that our nominated RStP value does not differ from the true value by more than 1\% for 30 to 300 MeV protons, with the largest error manifested in cortical bone. Further, the error is less than 2\% over the entire range of 10 to 300 MeV, and the residual energy of a proton exiting a patient is generally larger than both of these lower limits.

\begin{figure}[t]
\centering
\includegraphics[width=0.6\textwidth]{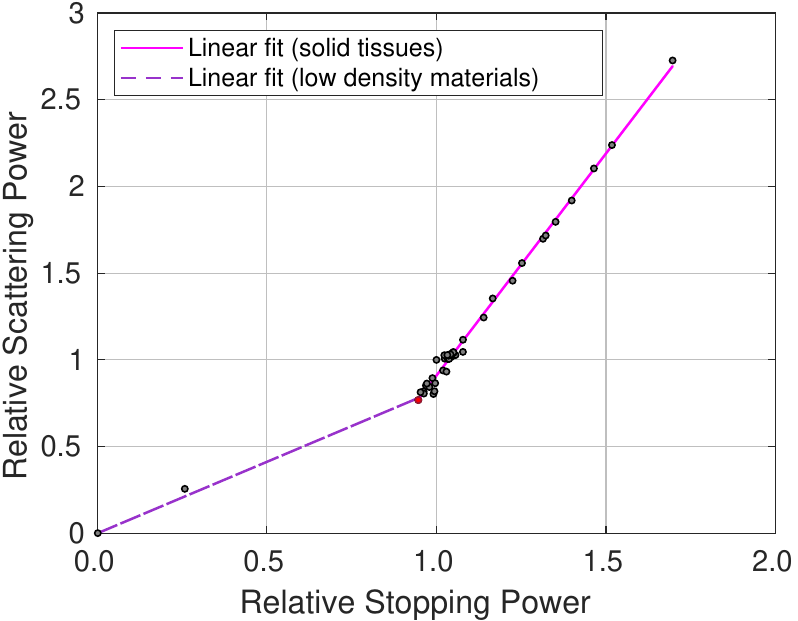}
\caption{Correlation between relative stopping and scattering powers ($\hat{S}$ and $\hat{T}$ respectively) for the materials listed in ICRU Report 46 \citep{ICRU46}. Linear least squares fits are shown for solid tissues and low density materials separately, with the intersection point (in red) defined at adipose tissue (\emph{Adipose Adult \#3} in \citep{ICRU46}). Parameters for the fits are detailed in \Tab{\ref{params}.}}\label{relSrelT}
\end{figure}

\subsection{Correlation of relative scattering and stopping powers in human tissues}\label{correlationsec}

As mentioned above, in an iterative pCT reconstruction algorithm, the RStP values in each voxel may be updated on successive iterations to build an image of the body. Implementation of a \emph{calibration curve} that determines RScP from RStP could provide an improvement to the convergence of the algorithm and the final accuracy of the image.

\begin{table}[b]
\centering
\caption{Parameters for the least squares fits, $\hat{S} = \hat{p}_1\hat{T}+\hat{p}_2$, on solid tissues and low density materials. $0 < r^2 \leq 1$ is the correlation coefficient (with $r^2 \approx 1$ indicating a good quality of fit for a sufficiently large data set). $N_\mathrm{s}$ is the number of samples in a group.}\label{params}
\footnotesize
\renewcommand{\arraystretch}{1.5}
\begin{tabular}{l r r r r}
 & $\hat{p}_1$ & $\hat{p}_2$ & $r^2$ & $N_\mathrm{s}$ \\
 \hline
 Solid tissues & 0.3905 & 0.6448 & 0.9945 & 45 \\
 Low density materials & 1.2127 & 0 & 0.9958 & 4
\end{tabular}
\end{table}

In \Fig{\ref{relSrelT}} we see strong evidence of a linear relationship between $\hat{T}$ and $\hat{S}$ for the solid tissues listed in ICRU report 46 \citep{ICRU46}. We use the adipose tissue with the lowest stopping power to define the intersection point (shown by the red marker) between the low density materials and solid tissues. A linear relationship is also expected for the low density materials as the chemical composition will be roughly equivalent and thus it is only the density increase in both parameters defining the correlation.

A linear least-squares model is applied to the data. In this method we assume
\begin{equation}
\hat{S} = p_1\hat{T} + p_2
\end{equation}\label{calibrationeq}
where $p_1, p_2 \in \mathbb{R}$ are constant parameters, given in \Tab{\ref{params}}, and $p_2$ is constrained to zero for low density materials.

\subsection{Monte Carlo simulations}\label{MCsec}

\begin{figure}[t]
\centering
\includegraphics[width=0.6\textwidth]{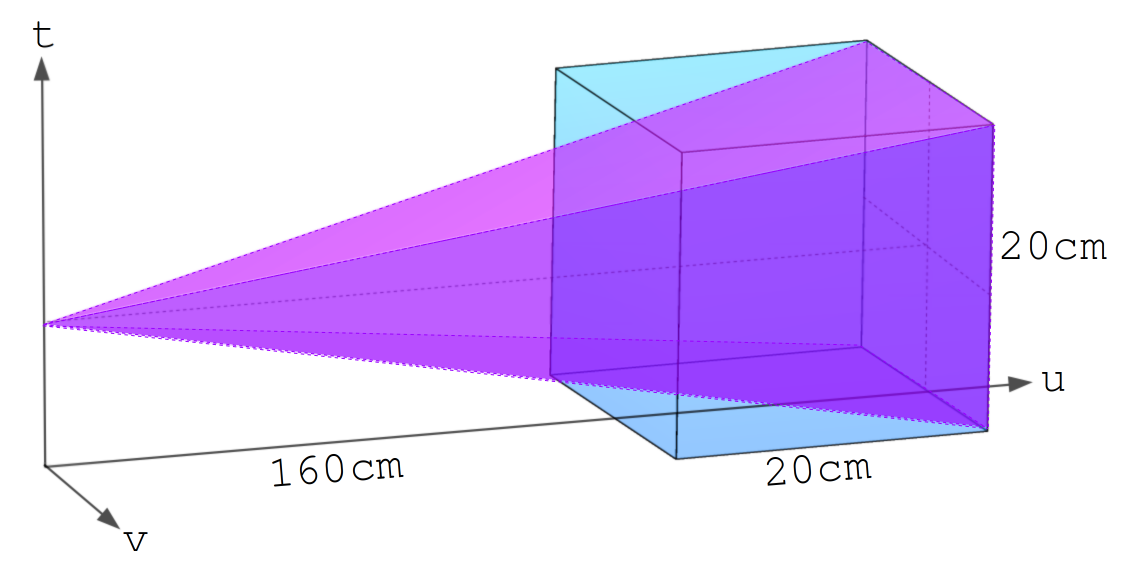}
\caption{Geometric setup in \emph{TOPAS} of a fan beam of protons incident on a 20cm cube of homogeneous water. The beam exactly spans the distal surface of the phantom.}\label{fanbeam}
\end{figure}

The performance of the inhomogeneous MLP formalism, MLP$_\mathrm{x}$, proposed in this paper was evaluated by comparison with proton tracks produced by Monte Carlo simulation using \emph{TOPAS}, version 3.1.2 using the standard environment and physics lists \citep{TOPAS}. MLP$_\mathrm{x}$ was tested against the homogeneous formalism, MLP$_\mathrm{H2O}$ by Schulte \emph{et al.} \citep{penfoldMLP} in three phantom setups, a homogeneous water cube, a water cube containing thick transverse slabs of bone, and a clinically relevant human head phantom. The effect on the performance of both algorithms as the nominal beam energy is decreased has also been investigated.

A monoenergetic planar fan beam of 200 MeV protons was delivered to a 20cm cube of homogeneous water such that the distance between the particle source and the surface of the phantom was 160 cm. The fan beam was parameterised such that the beam exactly spanned the exit face of the phantom, as shown in \Fig{\ref{fanbeam}}). Proton histories, including energy, position and direction were collected at 5 mm depth increments. These query depths should not be confused with the step-size in the Monte Carlo simulation itself. The step-size and included physics processes were unaltered from the default settings in TOPAS, which have been optimised for proton therapy applications \citep{Zacharatou08}. Increasing the sampling rate for the MLP calculation (at increments of smaller than 5 mm), and therefore decreasing the step-size in the Euler method, (24) and (25), and in the numerical integration, (4) to (6), did not yield any significant improvements in accuracy in the case of the slab phantom. 

The entire setup was placed in a vacuous world volume and two vacuum sensitive detectors were placed pre- and post-phantom to record the position, direction and energy of each proton before and after it traversed the cube. In an effort to retain only Gaussian-natured MCS events, data cuts were performed to remove tracks for which the total energy loss or projections of the relative exit angle onto both the $t$-$u$ and $u$-$v$ planes (refer to \Fig{\ref{fanbeam}}) differed from their mean values by more than a chosen number of standard deviations. Once the data cuts were performed, the MLP was only calculated for the projection of the path onto the $t$-$u$ plane, as the scattering in orthogonal directions form two independent probabilistic processes \citep{penfoldMLP}.

The above procedure was repeated for a 20 cm inhomogeneous cube consisting of 2 cm of water, 7 cm of cranium bone, 2 cm of cortical bone, another 7 cm of cranium bone and another 2 cm of water (in this order). Nominal beam energies of 230, 225, 220, 215 and 210 MeV were tested to investigate the robustness of MLP$_\mathrm{x}$ and MLP$_\mathrm{H2O}$ as the overall likelihood of scattering is increased as a result of decreasing the beam energy. Implementation of MLP${_\mathrm{H2O}}$ for beam energies other than 200 MeV requires recalculation of the fifth order polynomial for the energy factor, $\beta^{-2}p^{-2}$, as detailed by Schulte \emph{et al.} \citep{penfoldMLP}. The average proton energy was recorded at 5 mm increments in the 20 cm water cube for each nominal beam energy, which was then used to calculate the energy factors.

\begin{figure}[t]
\centering
\includegraphics[width=0.88\textwidth]{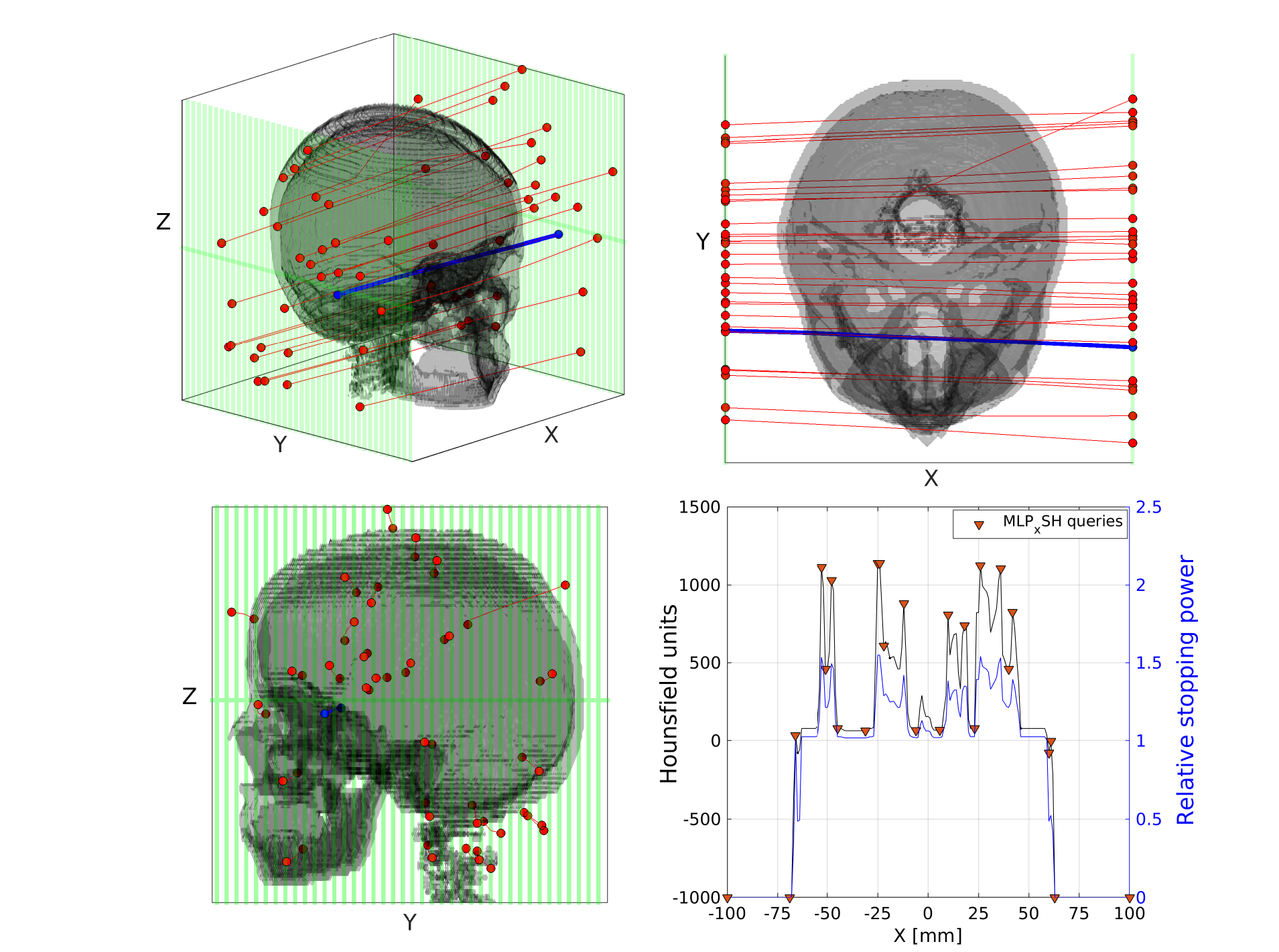}
\caption{Sample proton tracks (prior to statistical cuts) through a human head phantom (Computational Human Phantom Series, U.S. National Institutes of Health, National Cancer Institute, 2019) \citep{Lee2010,Lee2015}. Bottom right: Hounsfield units and RStP profiles for the particle track pictured in blue. An example of query points for implementing the MLP Spline-Hybrid method is shown. The depth coordinate $X$ here is equivalent to $u$, which is used elsewhere in this paper.}\label{headphantom}
\end{figure}

Finally, the slab phantom was replaced with a human head phantom to better resemble a clinical scenario. A 10-year-old male phantom (Computational Human Phantom Series, U.S. National Institutes of Health, National Cancer Institute, 2019) \citep{Lee2010,Lee2015} was imported into TOPAS as standard Digital Imaging and Communication in Medicine (DICOM) CT files using the internal GDCM package \citep{TOPAS}. The beam and detector setups were unchanged from the water phantom setup. A monoenergetic 200 MeV beam was delivered to the phantom and proton histories were inquired at 1 mm depth increments, due to the increased complexity of the geometry. The phantom spanned up to 143 mm in the beam direction. Each slice along the longitudinal axis is separated by 2.425 mm and each pixel in the transverse plane has dimensions of 0.99 mm by 0.99 mm. The full setup is shown in \Fig{\ref{headphantom}}, which includes sample proton tracks (prior to statistical cuts) for illustrative purposes. Further data cuts were performed to only include particle tracks that traversed the phantom.

Unlike the transverse slab geometry, for which there is only one material present at depth $u$, the material composition of a realistic phantom is voxelised and must be treated as such in the implementation of $\mathrm{MLP}_\mathrm{x}$ and $\mathrm{MLP}_\mathrm{x}$SH. For each proton registered at the prior and posterior detectors, the entry and exit positions and directions were used to fit a three-dimensional cubic spline. This approximate path was subsequently used to interpolate the RStP in the traversed voxel at depth $u$. The RStP was pre-calculated for each CT coordinate using the HU-RStP calibration of Schneider \emph{et al.} \citep{schneideretal}. The RScP was then calculated using (\ref{calibrationeq}), allowing the calculation of the inhomogeneous MLP. (Refer to Algorithm \ref{scatpowcalcalgo} and Algorithm \ref{inhomogMLPalgo} in the Appendix for more details on the implementation.) We note that while better accuracy may be attained by using the material composition directly in the calculation of RStP and RScP, our procedure has been designed to mimic the approximation of both values through the iterative reconstruction process, as we suggested in section \ref{correlationsec}.

Our application of the Spline-Hybrid method to iterative reconstruction of voxelised, non-slab geometries involves thresholding the current image estimate with three (or more) predefined regions corresponding to bone-like materials, tissue-like materials and air. An edge detection filter can be applied to the current iterate of the image based on the differences between RStP values in these regions, in much the same way that the outer patient boundary is detected on the first iteration in current pCT reconstruction methods \citep{penfoldMLP,Schultze2014}. For instance, the categories of solid tissues and low density materials defined in \Fig{\ref{relSrelT}} may be subdivided further to define a greater number of regions as the number of iterations increases. Here, our implementation of an edge-detection filter resulted in an average subsampling rate of 10\% of the available MLP$_\mathrm{x}$SH calculation points. An example of this subsampling is shown in \Fig{\ref{headphantom}} (bottom right).

When comparing our inhomogeneous formalism to the homogeneous water approach, a hull-detection method was used prior to the calculation of MLP$_\mathrm{H2O}$, as this is routinely performed in practice.

\section{Results}

\begin{table}[t]
\setlength{\tabcolsep}{12pt}
\renewcommand{\arraystretch}{1.2}
\small
\begin{center}
\caption{Percentage of proton tracks that escape the MLP $3\sigma$ envelope for the water and slab phantoms. This measure provides a point of comparison between the homogeneous and inhomogeneous MLP path predictions for different levels of data cuts in both relative exit angle and total energy loss.}\label{results}
\begin{tabular}{llrrrlllllllll}
                                                                                                                                       &                                                                  & \multicolumn{3}{c}{tracks outside 3$\sigma$ envelope (\%)}                                                                                                                                            \\ \cline{3-5}
                                                                                                                                       &                                                             \multicolumn{2}{r|}{\textbf{MLP}$_\mathrm{H2O}$} & \multicolumn{1}{c|}{\textbf{MLP}$_\mathrm{x}$} & \multicolumn{1}{l}{\textbf{MLP}$_\mathrm{x}$\textbf{SH}} \\ \cline{2-5}
\multicolumn{1}{l|}{\multirow{2}{*}{\begin{tabular}[c]{@{}l@{}}Water phantom \\ (200 MeV)\end{tabular}}} & No data cuts                                                     & \multicolumn{1}{r|}{2.47}                                         & \multicolumn{1}{r|}{2.44}                                                           & --           \\
\multicolumn{1}{l|}{}                                                                                                                  & 3$\sigma$ cuts                                                      & \multicolumn{1}{r|}{0.444}                                        & \multicolumn{1}{r|}{0.419}                                                          & --           \\
\multicolumn{1}{l|}{}                                                                                                                  & 2$\sigma$ cuts                                                      & \multicolumn{1}{r|}{0.251}                                        & \multicolumn{1}{r|}{0.214}                                                          & --           \\ \cline{1-5}
\multicolumn{1}{l|}{\multirow{5}{*}{Slab phantom}}                                                                                     & \begin{tabular}[c]{@{}l@{}}2$\sigma$ cuts \\ (210 MeV)\end{tabular} & \multicolumn{1}{r|}{9.63}                                         & \multicolumn{1}{r|}{0.223}                                                          & 0.208       \\
\multicolumn{1}{l|}{}                                                                                                                  & \begin{tabular}[c]{@{}l@{}}2$\sigma$ cuts\\ (215 MeV)\end{tabular}  & \multicolumn{1}{r|}{3.91}                                         & \multicolumn{1}{r|}{0.264}                                                          & 0.289       \\
\multicolumn{1}{l|}{}                                                                                                                  & \begin{tabular}[c]{@{}l@{}}2$\sigma$ cuts\\ (220 MeV)\end{tabular}  & \multicolumn{1}{r|}{2.34}                                         & \multicolumn{1}{r|}{0.183}                                                          & 0.204       \\
\multicolumn{1}{l|}{}                                                                                                                  & \begin{tabular}[c]{@{}l@{}}2$\sigma$ cuts \\ (225 MeV)\end{tabular} & \multicolumn{1}{r|}{1.75}                                         & \multicolumn{1}{r|}{0.250}                                                          & 0.254       \\
\multicolumn{1}{l|}{}                                                                                                                  & \begin{tabular}[c]{@{}l@{}}2$\sigma$ cuts\\ (230 MeV)\end{tabular}  & \multicolumn{1}{r|}{1.42}                                         & \multicolumn{1}{r|}{0.215}                                                       & 0.215       \\
\end{tabular}
\end{center}
\end{table}

\subsection{Water phantom}

\begin{figure}[t]
\centering
\includegraphics[width=0.6\textwidth]{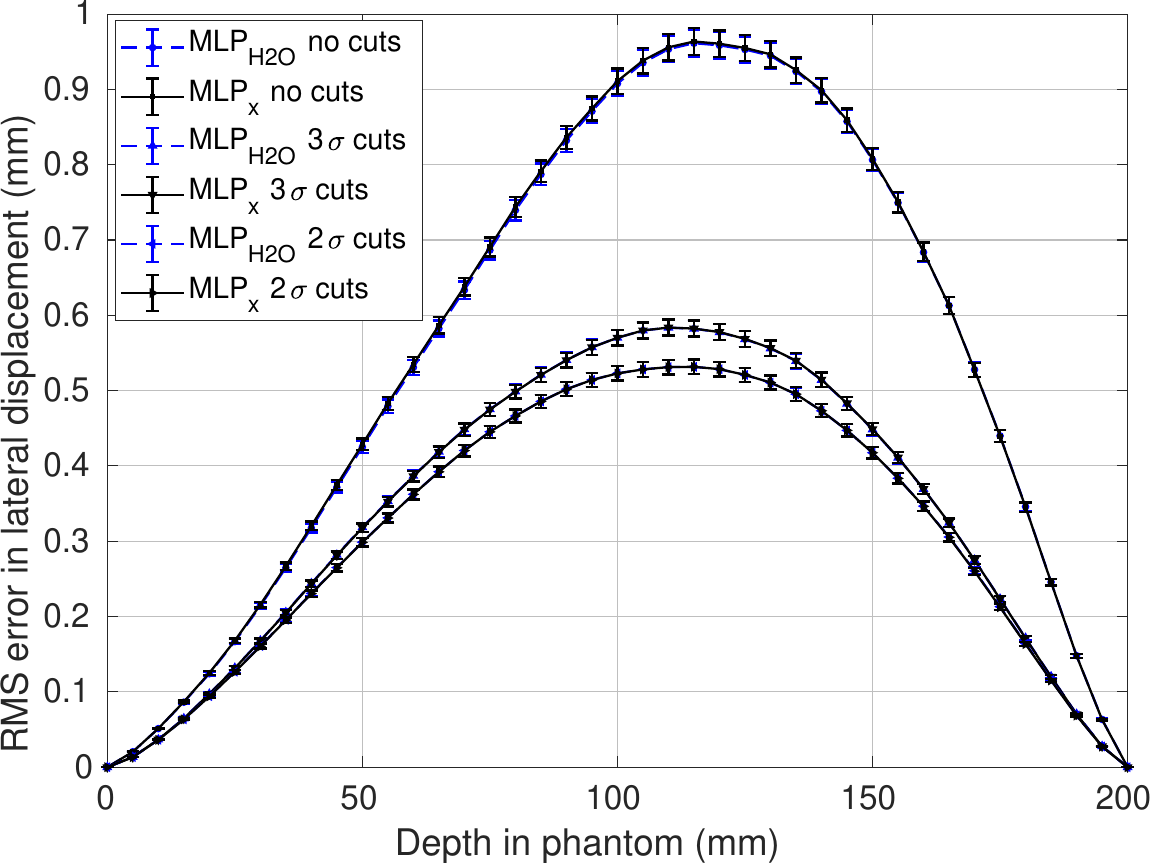}
\caption{Root mean square (RMS) error in lateral displacement in a 20 cm water cube over 3000 proton histories. MLP$_\mathrm{x}$ is compared to MLP$_\mathrm{H2O}$ for no data cuts, 2$\sigma$ cuts and 3$\sigma$ cuts in both relative exit angle and total energy loss.}\label{laterr_water}
\end{figure}

The error in the MLP as a function of depth, shown in \Fig{\ref{laterr_water}}, demonstrates that the inhomogeneous formalism performs as required, giving the same outcome in water as the homogeneous formalism when data cuts are applied to minimise the effects of elastic nuclear collisions and large-angle MCS. After $2\sigma$ cuts are performed on the relative exit angle and total energy loss, the maximum RMS error in lateral displacement is approximately 0.56 mm. A suitably accurate MLP estimation should give not only a good estimation of the trajectory, but also of the surrounding probability envelope. Indeed, \Tab{\ref{results}} shows that at least 99.7\% of particle tracks fall within the 3$\sigma$ envelope, as expected, for both formalisms.

\subsection{Slab phantom}

\begin{figure}[t]
\centering
\includegraphics[width=0.8\textwidth]{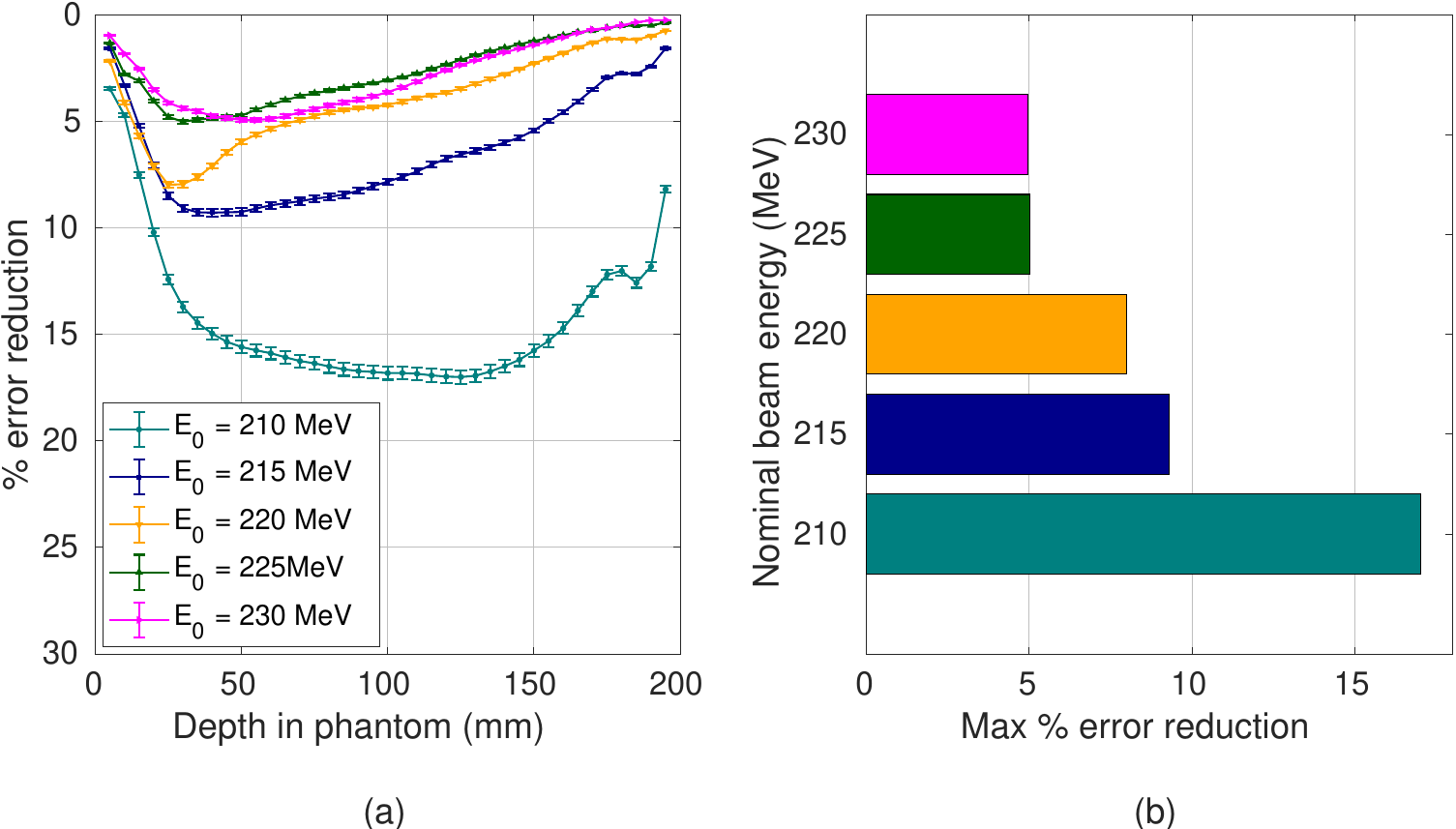}
\caption{(a) Percentage reduction in the root mean square (RMS) error in lateral position when MLP$_\mathrm{x}$SH is chosen over MLP$_\mathrm{H2O}$ for various nominal beam energies incident on the slab phantom. (b) Maximum percentage error reduction with each nominal beam energy.}\label{errorreductions}
\end{figure}

The inhomogeneous formalism when applied to 210 MeV protons traversing the slab phantom achieved a 17\% reduction in maximum RMS lateral position error when compared to the homogeneous formalism. This can be seen in \Fig{\ref{slabs}}(a) and \Fig{\ref{errorreductions}}. Both algorithms had a runtime of approximately 3 ms per particle track using \emph{MATLAB} (MATLAB R2017b, The MathWorks, Inc., Natick, Massachusetts, United States) on a standard \emph{Intel Core i7} processor. The spline-hybrid method, MLP$_\mathrm{x}$SH, regularly achieved very similar results up to 3 times faster on average. It can be seen in \Fig{\ref{errorreductions}} that at lower energies, at which scattering is more pronounced, MLP$_\mathrm{x}$SH retains greater accuracy than MLP$_\mathrm{H2O}$. After applying 2$\sigma$ statistical cuts, the percentage of tracks falling outside the 3$\sigma$ envelope, listed in \Tab{\ref{results}}, was consitently less than 0.3\% using the inhomogeneous formalism, indicating that the probabilistic path estimation is suitably accurate. On the contrary, significantly more tracks escaped the MLP$_\mathrm{H2O}$ envelope, with only 90.4\% being contained for the 210 MeV beam. An example proton track is shown in \Fig{\ref{slabs}}(c) along with all three MLP estimates. It is clear that MLP$_\mathrm{x}$ and MLP$_\mathrm{x}$SH give a larger and more right-skewed probability envelope surrounding the proton path when compared to MLP$_\mathrm{H2O}$ for this particular track.

\subsection{Human head phantom}

\begin{figure}[t]
\centering
\includegraphics[width=0.95\textwidth]{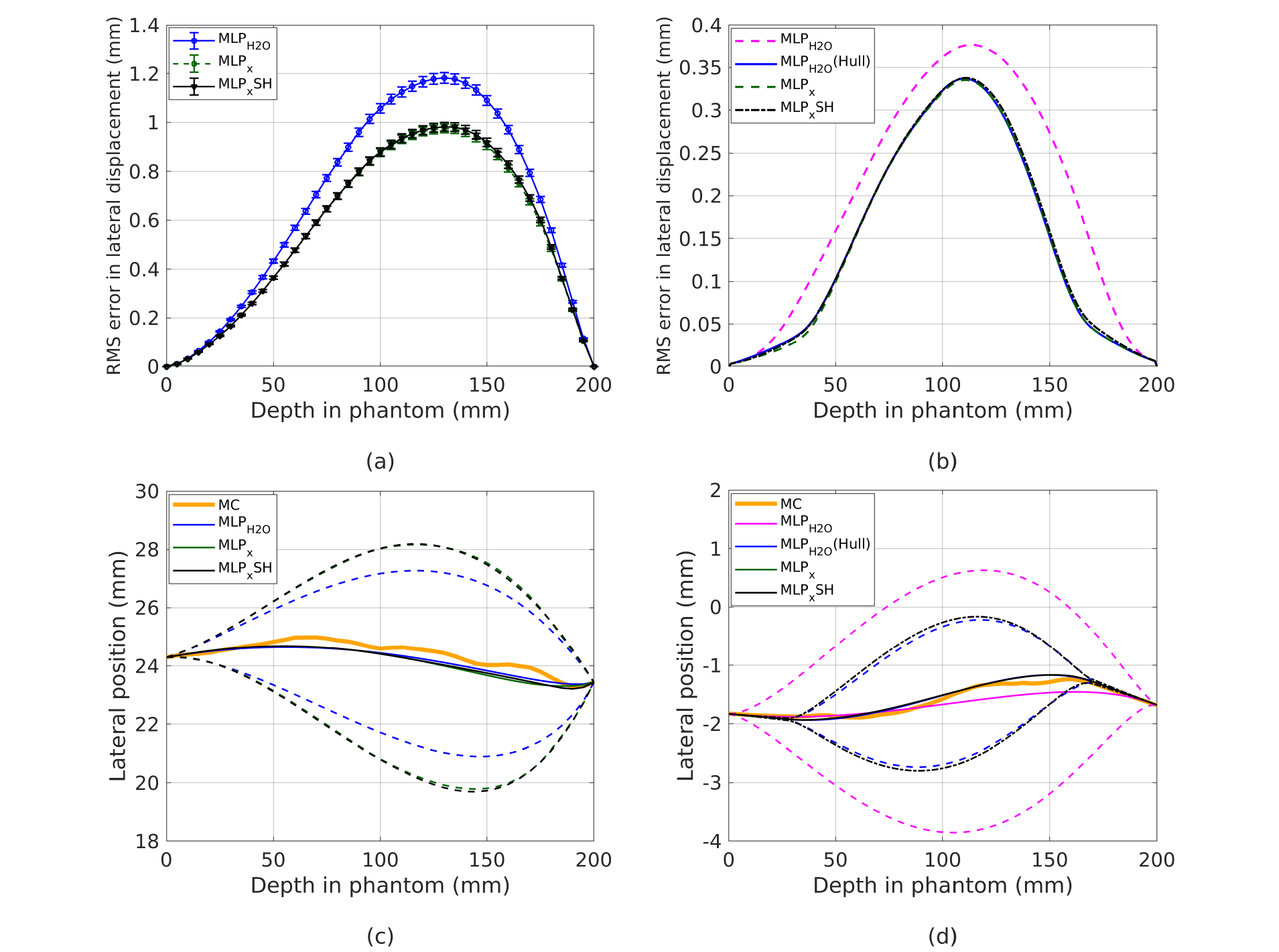}
\caption{(a,b) Root mean square (RMS) error in lateral displacement in (a) the slab Phantom over 3000 proton histories, and (b) the human head phantom over 10,000 proton histories. MLP$_\mathrm{H2O}$ is compared to MLP$_\mathrm{x}$ and MLP$_\mathrm{x}$SH after applying 2$\sigma$ cuts in both relative exit angle and total energy loss. In the case of the human head phantom, MLP$_\mathrm{H2O}$ has also been applied following a hull-detection step. (c,d) Examples of proton tracks in (c) the slab phantom, and (d) the human head phantom, showing the actual Monte Carlo (MC) path, the path as inferred by the different MLP methods, including their surrounding 3$\sigma$ probability envelopes. Note: error bars have been omitted for (b) as they are neglible.}\label{slabs}
\end{figure}

There was no improvement in accuracy by employing the inhomogeneous MLP formalism for the human head phantom when compared to the homogeneous method with a prior hull-detection, denoted MLP$_\mathrm{H2O}$(Hull). This is evident in \Fig{\ref{slabs}}(b). It can be seen in \Fig{\ref{slabs}}(d) that there is also neglibile difference between the surrounding envelopes of the inhomogeneous formalism and MLP$_\mathrm{H2O}$(Hull). Both MLP$_\mathrm{x}$ and MLP$_\mathrm{H2O}$(Hull) recorded runtimes consistently below 24 ms per track, however MLP$_\mathrm{x}$SH was able to complete the task in half this time.

\section{Discussion}\label{discussionsec}

A new inhomogeneous most likely path formalism using the matrix methods of Schulte \emph{et al.} \citep{penfoldMLP} has been proposed for use in iterative image reconstruction in proton computed tomography. Gottschalk's method \citep{gottschalkRadioProtons} has been used to calculate the relative scattering power (RScP) values of protons between 3 and 300 MeV for various human tissues. The relative stopping power (RStP) was calculated for this same energy range, showing negligible energy dependence (less than 1\% for energies above 30 MeV, and less than 2\% for those above 10 MeV). We must note, however, that the water equivalent thickness of the slab phantom is not much smaller than the range of 210 MeV protons. Consequently, an average residual energy of 15.7 MeV was recorded post-phantom and 19.7\% of protons had a residual energy of less than 10 MeV (after 2$\sigma$ data cuts were applied). No other beam energies resulted in a residual energy of less than 10 MeV, however 5.25\% of recorded protons from the 215 MeV beam had an energy of less than 30 MeV. It was necessary to determine to what degree the improvement in MLP accuracy was due to the inhomogeneous scattering moments, rather than simply the energy estimate via the Euler method given in \Eqn{\ref{eulerforward2}} and \Eqn{\ref{eulerbackward2}}. We therefore recalculated MLP$_\mathrm{x}$ (and MLP$_\mathrm{x}$SH) using a 5-th order polynomial estimate of the mean energy, akin to the energy factor used in MLP$_\mathrm{H2O}$ \citep{penfoldMLP}. There was no noticable difference (less than 0.3\%) in the RMS MLP error between these two methods, which leads us to conclude that the improvement in MLP accuracy was almost entirely due to the inclusion of inhomogeneity in the scattering moments.

A bi-linear correlation between RScP and RStP has been shown, which may be implemented in iterative pCT reconstruction as the RStP in each voxel is updated on successive iterations. Using the inhomogeneous formalism in a 20 cm cube of water, MLP$_\mathrm{x}$ gave the same path as MLP$_\mathrm{H2O}$ but with a slightly larger probability envelope when comparing estimates to Monte Carlo data obtained using \emph{TOPAS} \citep{TOPAS}. Note that the implementation of MLP$_\mathrm{H2O}$ requires calculation of the energy factor polynomial $\beta^{-2} p^{-2}$, and therefore simulation of protons traversing the required depth at the correct nominal beam energy. Many polynomial coefficients may be pre-calculated and stored, however this is not required when calculating MLP$_\mathrm{x}$.

In the inhomogeneous slab geometry consisting of large slabs of cortical and cranium bone, MLP$_\mathrm{x}$ achieved greater accuracy than MLP$_\mathrm{H2O}$ by up to 17\% at the lowest beam energy of 210 MeV. It was shown that the improvement in accuracy gained by using MLP$_\mathrm{x}$ over MLP$_\mathrm{H2O}$ increased as the beam energy was decreased and scattering in the phantom was more pronounced. This robustness in the inhomogeneous formalism has potential to be valuable in situations where lower energy beams are used. For instance, lower energy pencil beams are advantageous when aimed toward small volumes of matter near the body boundary, to ensure that the Bragg peak occurs within the absorbing detector post-patient. Furthermore, density resolution in pCT imaging has been shown to increase when using lower energy protons \citep{schultedensityres}.

MLP$_\mathrm{x}$ has in general outperformed MLP$_\mathrm{H2O}$ in cases of significant inhomogeneity. It is important to note that a suitably accurate MLP formalism should accurately infer both the MLP itself and the probability envelope surrounding it. After applying 2$\sigma$ data cuts, all inhomogeneous MLP estimates in this study contained at least 99.7\% of tracks within the 3$\sigma$ probability envelope, as expected. This was not the case when applying the water assumption to the slab geometry, as the number of tracks falling outside the envelope was significantly greater than 0.3\% for all nominal beam energies.

Throughout this paper we have referred to Gottschalk's method for calculating material-dependent scattering power, assuming locality \citep{gottschalkRadioProtons}. That is, the scattering power depends on only local properties of the material and the proton energy. In reality, one must consider the buildup of MCS events in material traversed prior to the point of calculation. Gottschalk defined a nonlocal extension to \Eqn{\ref{gottschalkscatpower2}} which differs by a simple logarithmic single scattering correction factor, $f_{dM}(E_0,E(u))$, depending only on the initial energy of the proton, $E_0$, and its energy at depth $u$. Namely,
\begin{equation}
T_{dM}(u) = f\left(E_0,E(u)\right)\frac{2\pi}{\alpha}\left( m_ec^2 \right)^2\left( \frac{\tau+1}{\tau+2} \right)^2\frac{1}{E^2(u)}\frac{1}{X_s}. \label{gottschalkscatpower3}
\end{equation}
We have adopted the subscript ``dM'' (for ``differential Moli\`ere'') from Gottschalk's work. Gottschalk showed that this nonlocal formula outperformed the differential Highland formula presented by Kanematsu \citep{Kanematsu2008}, off which the inhomogeneous MLP proposed by Collins-Fekete \emph{et al.} \citep{Collins2017} is based. It was also shown to be more easily implemented in mixed slab geometries as the non-locality parameter is not material dependent. Contrastingly, Kanematsu's method requires numerical integration over a material-dependent quantitiy, and its accuracy is thus subject to one's confidence in the material composition preceeding the point of interest, and the chosen step-size for numerical integration. It was for these reasons that Gottschalk's method was chosen for calculating the scattering power in our proposed MLP formalism. 

Using \Eqn{\ref{gottschalkscatpower3}} we still obtain the same expression \Eqn{\ref{relscatpower}} for the RScP, however the scattering power in water differs in \Eqn{\ref{totalscatpower}}. We found negligible (less than 1\%) difference in terms of MLP RMS error and the percentage of paths falling outside the 3$\sigma$ envelope when implementing the non-local correction to the scattering power. Thus, the local version \Eqn{\ref{gottschalkscatpower2}} was chosen for simplicity. This is due primarily to the fact that the proton entry and exit information are known within the Bayesian MLP formalism. It was found that, while the non-local correction is substantially large at energies that are either very low or very close to the input energy, the depths at which a proton exhibits these energies are near either the point of entry or exit. Elsewhere in the phantom, the effect of the correction is very close to null.

In terms of designing a practical pCT scanner, the intended energy range of the protons determines the required water equivalent thickness (WET) of the residual range telescope (RRT), or calorimeter. For example, a residual energy of 115 MeV post-patient will result in a WET of approximately 10 cm. In initial designs, a crystal calorimeter was prescribed a depth appropriate for stopping protons with an initial energy of 200 MeV during a head scan \citep{Bashkirov2016}. In a solid-state pCT system design by Esposito \emph{et al.} \citep{pravda}, an RRT with a WET of 55 mm was used, corresponding to a maximum residual energy of approximately 80 MeV. Increasing the incident proton energy, in addition to requiring a physically larger RRT, can lead to increased uncertainty in the measured range \citep{Price2015}. Collins-Fekete \emph{et al.} \citep{Collins2017} included only simulations of 330 MeV protons with a head, thorax and abdominal phantom. While this proton energy is favourable in terms of reduced MCS, accurate residual energy detection, particularly for a head-sized object, would be challenging with current design concepts. We have assessed our inhomogeneous MLP algorithm with proton energies more suited to residual range detector designs for proton imaging of head-like objects.

The detectors in our setup were placed 1 cm from the scoring geometry on either side. A linear projection from the detector to the phantom in the direction of motion was made to determine the entry and exit positions at the scoring boundary, as negligible scatter contribution is expected as the proton traverses air.

It is important to note that large volumes of high density material such as those in the slab phantom will rarely be encountered in clinical practice, and so a gain in lateral position accuracy alone from MLP$_\mathrm{x}$ may be insignificant in many cases. However, as the variance in scattering probability is increased in higher density materials, the width and skewness of the probability envelope surrounding the MLP within the inhomogeneous formalism are variable and better representative of the confidence in the expected lateral position. This quality has the potential to be useful in probabilistic reconstruction (see, for example, \citep{wangmackietome}) of high density materials, if the volume is sufficiently large or the beam energy is sufficiently low. A right-skewed envelope is evident for the slab geometry tested in this work, which can be explained by the fact that protons, on entry, have larger momentum and therefore lower scattering power. Scattering is more pronounced toward the distal part of the phantom where the particle has lower kinetic energy.

In this study, we have only considered inhomogeneous slabs transverse to the initial beam direction which extend to the boundaries of the phantom. It would be beneficial to explore the effect of structures both finite in size and parallel to the beam direction on the accuracy of the inhomogeneous MLP. In such a setup, discontinuities in the scattering of protons at the boundary could be expected to introduce a systematic error in the MLP under the homogeneous water assumption \citep{Arbor2015}. The reader is referred to the recent investigation by Khellaf \emph{et al.} \citep{Khellaf2019} on the effects of transverse tissue heterogeneities on the accuracy of the conventional water MLP formalism by Schulte \emph{et al} \citep{penfoldMLP}.

Clinical applicability of our formalism was investigated by applying these methods to a human head phantom. There was no noticeable increase in path estimation accuracy using the inhomogeneous formalism a beam of monoenergetic 200 MeV protons, when compared to the standard water assumption with a prior hull-detection step. Our results indicate that in the majority of medical scenarios, the assumption of a water scattering environment is valid.

The MLP calculation is the most computationally intensive aspect of iterative pCT reconstruction. Both MLP$_\mathrm{x}$ and MLP$_\mathrm{H2O}$(Hull) performed similarly in terms of computational time, however MLP$_\mathrm{x}$SH was twice as fast on average, with no appreciable loss in accuracy. The speed of the algorithm may be improved further through redefining the subsampling technique, using an alternative hull-detection technique, or interpolating the RStP voxel values using a straight line path instead of a cubic spline. However it should be noted that techniques such as parallelisation are able to greatly improve the computational efficiency of the MLP calculation under the assumption of a water scattering environment, and a prior hull-detection step significantly reduces the number of calculation points. Our proposed inhomogeneous formalism has the potential advantage of improving convergence for air inside the body, and while a hull is suited well to a convex patient exterior, our method may be able to handle non-convexities on the exterior better than current methods such a filtered back-projection.

\section{Conclusions}

A new formalism for calculating the most likely path of protons through inhomogeneous matter has been proposed, based on the compact matrix methods of Schulte \emph{et al.} \citep{penfoldMLP} and the scattering power formulae of Gottschalk \citep{gottschalkRadioProtons}. The inhomogeneous formalism was shown to predict Monte Carlo proton paths to within 1.0 mm on average for beams ranging from 230 MeV down to 210 MeV in a 20 cm phantom consisting of thick slabs of cortical and cranium bone. The improvement in accuracy was most noticeable at lower energies, ranging from 5\% maximum RMS error in the MLP for a 230 MeV beam to 17\% for 210 MeV, when compared to their corresponding Monte Carlo tracks. Implementation of a new MLP-Spline-Hybrid method offered reduced computation time while suffering negligible loss of accuracy. There was no accuracy improvement for simulated 200 MeV protons through a more clinically relevant human head phantom, suggesting that a water scattering envrionment can be assumed in most clinical cases. However, in certain cases of significant heterogeneity the proposed algorithm may improve proton path estimation.

\section{Acknowledgments}

M. Brooke is supported by a John Monash Scholarship and a Clarendon Fund Scholarship. This work was supported by Cancer Research UK grant number C2195/A25197, through a CRUK Oxford Centre DPhil Prize Studentship. The authors acknowledge C. Lee and the National Cancer Institute for providing data sets from the Computational Human Phantom Series (U.S. National Institutes of Health, National Cancer Institute, 2019).

\section{Declarations of interest}

None.

\section{ORCID iDs}

\noindent Mark Brooke: 0000-0002-2262-7205\\
Scott Penfold: 0000-0002-3422-9108

\newpage
\section*{Appendix: implementation of inhomogeneous most-likely-path algorithm}

Algorithm \ref{scatpowcalcalgo} followed by Algorithm \ref{inhomogMLPalgo} provides one method of implementing the inhomogeneous MLP formalism proposed in this work. Relative scattering powers and relative stopping powers for materials of interest may be calculated using the methods outlined in sections \ref{scatteringpowersection} and \ref{stoppingpowersection}, respectively. A vector of ordered discrete depths is passed to the MLP calculation function, along with corresponding vectors of RStP and RScP values at each depth, and the entry and exit information of the proton. Algorithm \ref{scatpowcalcalgo} calculates the material-dependent scattering power, which is then used by Algorithm \ref{inhomogMLPalgo} to calculate the most likely lateral position at each depth and the associated standard deviation in this position. The scattering moments in (\ref{scatmomentsSigma_1}) to (\ref{scatmomentsSigma_3}) are calculated using the trapezoidal rule for numerical integration, however alternate numerical integration methods are possible. The step-size for the numerical integration or for the Euler method can also be increased, for improving calculation time, or decreased, for improving accuracy. Experimentation showed insignificant accuracy improvement from using step-sizes smaller than 5mm for the inhomogeneous slab phantom in this work. Therefore, both algorithms use a step-size that is the same as the depth sampling rate.

\begin{algorithm}
\footnotesize
\DontPrintSemicolon
\KwIn{Discrete depth vector $\mathbf{u} = (u_j)_{j=0,...,N}$ (in units of cm) with vectors of relative stopping power, $\mathbf{RStP}$, and relative scattering power, $\mathbf{RScP}$. Entry energy $E_\mathrm{in}$ (MeV) and exit energy $E_\mathrm{out}$ (MeV)}
\KwOut{Material-dependent scattering power $\mathbf{T}$ vector corresponding to each depth.}\BlankLine
\BlankLine
Initialise energy vectors for forward ($\mathbf{E}^F$) and backward ($\mathbf{E}^B$) Euler methods as empty vectors of length $N+1$.\;
\BlankLine
\tcc{Set entry and exit energies:}
$\mathbf{E}^F(0)=\mathbf{E}^B(0)=E_\mathrm{in}$\;
$\mathbf{E}^F(N)=\mathbf{E}^B(N)=E_\mathrm{out}$\;
\BlankLine
\tcc{Forward Euler method:}
Initialise index $k = 1$\;
\While{$k \leq N-1$}{
	Calculate stopping power in water, $S_w$, at energy $\mathbf{E}^F(k-1)$ using (\ref{bethebloch}).\;
	$\mathbf{E}^F(k) \gets \mathbf{E}^F(k-1) - [u(k)-u(k-1)]. \mathbf{RStP}(k-1). S_w$\;
	$k \gets k + 1$\;
}
\BlankLine
\tcc{Backward Euler method:}
$k \gets N-1$\;
\While{$k \geq 1$}{
	Calculate stopping power in water, $S_w$, at energy $\mathbf{E}^B(k+1)$ using (\ref{bethebloch}).\;
	$\mathbf{E}^B(k) \gets \mathbf{E}^B(k+1) + [u(k+1)-u(k)]. \mathbf{RStP}(k+1). S_w$\;
	$k \gets k - 1$\;
}
\BlankLine
\tcc{Calculate the weighted energies:}
Initialise weighted energy vector $\mathbf{E}$ as empty vector of length $N+1$.\;
$k \gets 1$\;
\While{$k \leq N-1$}{
	$\mathbf{E}(k) \gets (N-k) \mathbf{E}^F(k)/N + k \mathbf{E}^B(k)$/N\;
	$k \gets k + 1$\;
}
Calculate the scattering power in water, $\mathbf{T}_w$, at all energies in $\mathbf{E}$ using (\ref{scatpower1})
\BlankLine
\tcc{Calculate the material-dependent scattering power $\mathbf{T}$ using (\ref{totalscatpower}).}
$\mathbf{T} = \mathbf{RScP} \otimes \mathbf{T}_w$\;
\Return{$\mathbf{T}$}\;
\caption{{\sc scattering power calculation along path}}
\label{scatpowcalcalgo}
\end{algorithm}

\begin{algorithm}
\footnotesize
\DontPrintSemicolon
\KwIn{Discrete depth vector $\mathbf{u} = (u_j)_{j=0,...,N}$ (cm), vector of material-dependent scattering power values at each depth $\mathbf{T}$, lateral entry position $t_\mathrm{in}$ (cm), exit position, $t_\mathrm{out}$ (cm), entry direction $\theta_\mathrm{in}$ (rad) and exit direction $\theta_\mathrm{out}$ (rad)}
\KwOut{Most likely position vector $\mathbf{t}_\mathrm{MLP}$ and standard deviation $\hat\mathbf{\sigma}_t$ vector corresponding to each depth. Both have units of cm.}\BlankLine
Calculate the scattering power in water, $\mathbf{T}_w$, at all energies in $\mathbf{E}$ using (\ref{scatpower1})
\BlankLine
\tcc{Calculate the most-likely-path:}
Set $\mathbf{y}_\mathrm{in} = (t_\mathrm{in},\theta_\mathrm{in})^T$\;
Set $\mathbf{y}_\mathrm{out} = (t_\mathrm{out},\theta_\mathrm{out})^T$\;
Initialise change of basis matrices $\mathbf{R}_0$ and $\mathbf{R}_1$ to $2\times 2$ identity matrices.\;
\BlankLine
$k \gets 1$\;
\While{$k \leq N-1$}{
	\tcc{Note, we use the convention of matrix row/column indices starting from 0.}
	$\mathbf{R}_0(0,1) \gets u(k)-u(0)$\;
	$\mathbf{R}_1(0,1) \gets u(N)-u(k)$\;
	\BlankLine
	\tcc{Calculate the scattering moments in (\ref{scatmomentsSigma_1}) to (\ref{scatmomentsSigma_3}) using the trapezoidal numerical integration method}
	Initialise covariance matrices $\mathbf{\Sigma}_1$ and $\mathbf{\Sigma}_2$ to $2\times 2$ zero matrices.\;
	\For{$j=1$ to $k$}{
	$\mathbf{\Sigma}_1(1,1) \gets \mathbf{\Sigma}_1(1,1) + 0.5[\mathbf{u}(j)-\mathbf{u}(j-1)][\mathbf{T}(j-1) + \mathbf{T}(j)]$\;
	\BlankLine
	$\mathbf{\Sigma}_1(0,0) \gets \mathbf{\Sigma}_1(0,0) + 0.5[\mathbf{u}(j)-\mathbf{u}(j-1)][(\mathbf{u}(k)-\mathbf{u}(j-1))^2\mathbf{T}(j-1) + (\mathbf{u}(k)-\mathbf{u}(j))^2\mathbf{T}(j)]$\;
	\BlankLine
	$\mathbf{\Sigma}_1(0,1) \gets \mathbf{\Sigma}_1(0,1) + 0.5[\mathbf{u}(j)-\mathbf{u}(j-1)][(\mathbf{u}(k)-\mathbf{u}(j-1)))\mathbf{T}(j-1) + (\mathbf{u}(k)-\mathbf{u}(j))\mathbf{T}(j)]$\;
	}
	$\mathbf{\Sigma}_1(1,0) \gets \mathbf{\Sigma}_1(0,1)$\;
	\For{$j=k$ to $N$}{
	$\mathbf{\Sigma}_2(1,1) \gets \mathbf{\Sigma}_2(1,1) + 0.5[\mathbf{u}(j)-\mathbf{u}(j-1)][\mathbf{T}(j-1) + \mathbf{T}(j)]$\;
	\BlankLine
	$\mathbf{\Sigma}_2(0,0) \gets \mathbf{\Sigma}_2(0,0) + 0.5[\mathbf{u}(j)-\mathbf{u}(j-1)][(\mathbf{u}(N)-\mathbf{u}(j-1))^2\mathbf{T}(j-1) + (\mathbf{u}(N)-\mathbf{u}(j))^2\mathbf{T}(j)]$\;
	\BlankLine
	$\mathbf{\Sigma}_2(0,1) \gets \mathbf{\Sigma}_2(0,1) + 0.5[\mathbf{u}(j)-\mathbf{u}(j-1)][(\mathbf{u}(N)-\mathbf{u}(j-1))\mathbf{T}(j-1) + (\mathbf{u}(N)-\mathbf{u}(j))\mathbf{T}(j)]$\;
	}
	$\mathbf{\Sigma}_2(1,0) \gets \mathbf{\Sigma}_2(0,1)$\;
	\BlankLine
	Calculate $\mathbf{y}_\mathrm{MLP}$ using (\ref{mlp}).\;
	\tcc{Extract the first element (position) and ignore the angle.}
	$\mathbf{t}_\mathrm{MLP}(k) \gets \mathbf{y}_\mathrm{MLP}(0,0)$
	\BlankLine
	Calculate the error matrix $\mathbf{M}$ using (\ref{errormatrixexact}).\;
	\tcc{Extract the element in the first row and first column.}
	$\hat\mathbf{\sigma}_t(k) = \mathbf{M}(0,0)$\;
	\BlankLine
	$k \gets k + 1$\;
}
\Return{$\mathbf{t}_\mathrm{MLP}$; $\hat\mathbf{\sigma}_t$}\;
\caption{{\sc inhomogeneous most-likely-path calculation}}
\label{inhomogMLPalgo}
\end{algorithm}

\end{document}